\documentclass[aps,superscriptaddress]{revtex4}
\usepackage{amssymb,amsmath,epsfig}
\usepackage{subfigure}
\usepackage[colorlinks=true, pdfstartview=FitV, linkcolor=blue, citecolor=red, urlcolor=magenta, breaklinks=true]{hyperref}
\begin{document}
\title{Absorption and scattering of a self-dual black hole}

\author{M. A. Anacleto}\email{anacleto@df.ufcg.edu.br}
\affiliation{Departamento de F\'{\i}sica, Universidade Federal de Campina Grande
Caixa Postal 10071, 58429-900 Campina Grande, Para\'{\i}ba, Brazil}

\author{F. A. Brito}\email{fabrito@df.ufcg.edu.br}
\affiliation{Departamento de F\'{\i}sica, Universidade Federal de Campina Grande
Caixa Postal 10071, 58429-900 Campina Grande, Para\'{\i}ba, Brazil}
\affiliation{Departamento de F\'isica, Universidade Federal da Para\'iba, 
Caixa Postal 5008, 58051-970 Jo\~ao Pessoa, Para\'iba, Brazil}
 
\author{J. A. V. Campos}\email{joseandrecampos@gmail.com}
\affiliation{Departamento de F\'isica, Universidade Federal da Para\'iba, 
Caixa Postal 5008, 58051-970 Jo\~ao Pessoa, Para\'iba, Brazil}

\author{E. Passos}\email{passos@df.ufcg.edu.br}
\affiliation{Departamento de F\'{\i}sica, Universidade Federal de Campina Grande
Caixa Postal 10071, 58429-900 Campina Grande, Para\'{\i}ba, Brazil}

\begin{abstract} 
In this paper we aim to investigate the process of massless scalar wave scattering due to a self-dual black hole through the partial wave method.
We calculate the phase shift analytically at the low energy limit and we show that the dominant term of the differential cross section at the small angle limit is modified by the presence of parameters related to the polymeric function and minimum area of a self-dual black hole.
We also find that the result for the absorption cross section is given by the event horizon area of the self-dual black hole at the low frequency limit.
We also show that, contrarily to the case of a Schwarzschild black hole, the differential scattering/absorption cross section of a self-dual black hole is nonzero at the zero mass limit.
In addition, we verify these results by numerically solving the radial equation for arbitrary frequencies.

\end{abstract}

\maketitle
\pretolerance10000

\section{Introduction}
The so-called self-dual black hole corresponds to a simplified model that is obtained through 
a semiclassical analysis of loop quantum gravity~\cite{Ashtekar:2004eh,Han:2005km,Thiemann:1992jj,Ashtekar:2008zu,Bojowald:2003uh}.
Loop quantum gravity is a quantum geometric theory constructed for the purpose of reconciling general relativity and quantum mechanics on the Planck scale.
This theory is derived from the canonical quantization procedure of the Einstein equations obtained in terms of the Ashtekar variables~\cite{Ashtekar:1986yd}. 
The metric of the self-dual black hole with quantum gravity corrections has been found in~\cite{Modesto:2008im,Modesto:2009ve}. This metric is characterized by its dependence on parameters $ P $ and $ a_0 $, where $ P $ is referred to as the polymeric parameter and $ a_0 $ as a minimum area in loop quantum gravity. It also features an event horizon and a  Cauchy horizon.
In addition, the condition of self-duality has the property of removing the singularity and replace it with an asymptotically flat region.
The Schwarzschild black hole solution is recovered in the limit as $ P $ and $ a_0 $ go to zero. An analysis on the evaporation of self-dual black holes has been performed in~\cite{Hossenfelder:2009fc,Hossenfelder:2012tc,Moulin:2018uap,Alesci:2011wn}. 
By employing the tunneling formalism via the Hamilton-Jacobi method the thermodynamics of a self-dual black hole has been analyzed in~\cite{Silva:2012mt}.
In addition, the authors in~\cite{Anacleto:2015mma}, by applying this same method investigated the effect of the generalized uncertainty principle on the thermodynamic properties of the self-dual black hole.
Studies on quasinormal modes of self-dual black holes using the WKB approximation have been performed in~\cite{Santos:2015gja,Cruz:2015bcj}.
In~\cite{Sahu:2015dea} the authors have discussed the gravitational lensing effect of a self-dual black hole.

In the present work we have the main purpose of exploring the effect of quantum gravity corrections that contribute 
to the process of massless scalar wave scattering by a self-dual black hole.
In this analysis we will make use of the partial wave method to calculate the absorption and differential cross section. 
For this we will apply the technique implemented in~\cite{Anacleto:2017kmg} in order to determine the phase shift analytically at the low frequency limit ($m\omega\ll 1$). 
This technique has also been considered in~\cite{Anacleto:2019tdj} to examine the problem of scalar wave scattering by a noncommutative black hole.
The partial wave approach has been widely applied in the black hole scattering process by considering 
various types of metrics~\cite{Futterman1988,Matzner1977,Westervelt1971,Peters1976,Sanchez1976,Logi1977,
Doram2002,Dolan:2007ut,Crispino:2009ki,Churilov1974,Gibbons1975,Page1976,Churilov1973,Moura:2011rr,
Jung2004,Doran2005,Dolanprd2006,Castineiras2007,Benone:2014qaa,Marinho:2016ixt,Das:1996we,
Macedo:2016yyo,deOliveira:2018kcq,Hai:2013ara}.
This approach has also been applied to exploit scalar wave scattering by acoustic black holes~\cite{Crispino:2007zz,Dolan:2009zza,Oliveira:2010zzb,Dolan:2011zza,Dolan:2012yc,ABP2012-1,Anacleto:2015mta,Anacleto:2018acl} and also in~\cite{Brito2015} the differential cross section of a noncommutative BTZ black hole has been obtained. 
In our analysis we have initially calculated the phase shift analytically and then computed the differential cross section and the absorption. We find that the results obtained for the absorption/differential cross section have their values increased when we vary the values of the polymeric parameter $ P $.
In addition, we show that at the mass limit going to zero the absorption and differential cross section tends to nonzero values that are proportional to the minimum area $ a_0 $ of the self-dual black hole.
Furthermore, we also have extended our computations for high energy regime by numerically solving the radial equation for arbitrary frequencies. 
We show that the numerical results have a good agreement with the results obtained analytically in the low frequency limit.

The paper is organized as follows. In Sec.~\ref{sc2} we derive the phase shift and calculate the differential scattering/absorption cross
section for a self-dual black hole by considering analytical and numerical analysis. In Sec.~\ref{conc} 
we make our final considerations. 

\section{Self-Dual Black Hole}
\label{sc2}
In this section we introduce the self-dual black hole for the purpose of determining the differential cross section and absorption for this model. 
Thus, we adopt the partial wave method to calculate the phase shift at the low energy limit. 
The self-dual black hole is described by the following line element
\begin{eqnarray}
\label{metrsd}
ds^2=F(r)dt^2-\frac{dr^2}{N(r)}-\rho^2(r)\left(d\theta^2 + \sin^2\theta d\phi^2 \right).
\end{eqnarray}
The functions $ F(r) $, $ N(r) $ and $ \rho(r) $ are given respectively by
\begin{eqnarray}
F(r) &=& \frac{(r-r_{+})(r-r_{-})(r+r_{*})^2}{r^{4}+a_{0}^{2}} ,
\\
N(r) &=& \frac{(r-r_{+})(r-r_{-})r^{4}}{(r+r_{*})^{2}(r^{4}+a_{0}^{2})} ,
\\
 \rho(r) &=& r\sqrt{1 + \frac{a_{0}^{2}}{r^{4}}},
\end{eqnarray}
where $ \rho $ denotes a physical radial coordinate.
Here the event horizon, $ r_{+} $, and the Cauchy horizon, $ r_{-} $, read 
\begin{eqnarray}
r_{+} = {2m}, \quad \quad r_{-} = 2m P^{2}.
\end{eqnarray}
We also define 
\begin{eqnarray}
r_{*} = \sqrt{r_{+}r_{-}} = 2mP,
\end{eqnarray} 
where $P$  is the polymeric function given by
\begin{equation}
P = \frac{\sqrt{1+\epsilon^{2}} - 1}{\sqrt{1+\epsilon^{2}} +1} , 
\end{equation}
with $ \epsilon = \gamma\delta  $, where $ \gamma $ is the Barbero-Immirzi parameter and $ \delta $ is the polymeric
parameter. In addition, we have  
\begin{eqnarray}
a_{0} = A_{min}/8\pi, 
\end{eqnarray}
which is the area gap and $ A_{min} $ is a minimum area in loop quantum gravity. 

Finally, the value of $ \rho $ that is related to the event horizon $ r_{+} $ is given by the following relation:
\begin{eqnarray}
\rho_{h}=\rho(r_{+}) &=& r_{+}\sqrt{1 + \frac{a_{0}^{2}}{r_{+}^{4}}}=2m\sqrt{1 + \frac{a_{0}^{2}}{(2m)^{4}}}.
\end{eqnarray}

\subsection{Absorption and Differential Scattering Cross Section}
We now consider the case of the massless scalar field equation to describe the scattered wave in the background (\ref{metrsd}),  given by 
\begin{eqnarray}
\dfrac{1}{\sqrt{-g}}\partial_{\mu}\Big(\sqrt{-g}g^{\mu\nu}\partial_{\nu}\Psi\Big)=0 .
\end{eqnarray}
We shall now apply a separation of variables into the equation above
\begin{eqnarray}
\Psi_{\omega l m}({\bf r},t)=\frac{{\cal R}_{\omega l}(r)}{r}Y_{lm}(\theta,\phi)e^{-i\omega t},
\end{eqnarray}
where $Y_{lm}(\theta,\phi)  $ are the spherical harmonics and $ \omega $ is the frequency.

Thus, we obtain the following  radial equation for $ {\cal R}_{\omega l}(r) $  
\begin{eqnarray}
\label{eqrad}
\Lambda(r)\dfrac{d}{dr}\left(\Lambda(r)\dfrac{d{\cal R}_{\omega l}(r)}{dr} \right) +\left[ \omega^2 -V_{eff} \right]{\cal R}_{\omega l}(r)=0,
\end{eqnarray}
for $ \Lambda(r)=\sqrt{F(r)N(r)} $ and $ V_{eff} $ being the effective potential, given by
\begin{eqnarray}
V_{eff}=\frac{\Lambda(r)}{\rho(r)}\left[\rho^{\prime}(r)\frac{d\Lambda(r)}{dr}+\Lambda(r)\rho^{\prime\prime}(r)\right]
+\frac{F(r)l(l+1)}{\rho^2(r)},
\end{eqnarray}
with
\begin{eqnarray}
\rho^{\prime}(r)=\frac{\rho(r)}{r}\left(1-\frac{2a^2_0}{\rho^2(r)r^2}  \right),
&&
\rho^{\prime\prime}(r)=\frac{6a^2_0}{\rho(r) r^4}\left(1-\frac{2a^2_0}{3\rho^2(r)r^2}  \right).
\end{eqnarray}
Next, to write equation (\ref{eqrad}) in the form of a Schroedinger-type equation, we introduce the change of variables,
$ \chi(r)=\Lambda^{1/2}(r){\cal R}(r) $, so we have
\begin{eqnarray}
\label{eqradpsi}
\dfrac{d^2\chi(r)}{dr^2}+V(r) \chi(r) = 0,
\end{eqnarray}
where
\begin{eqnarray}
\label{poteff}
V(r)=\dfrac{[\Lambda'(r)]^2}{4 \Lambda^2(r)} - \dfrac{\Lambda''(r)}{2\Lambda(r)} + \dfrac{\omega^2}{\Lambda^2(r)} - 
\dfrac{V_{eff}}{\Lambda^2(r)}.
\end{eqnarray}
Now,  by applying a power series expansion in $1/r$ for the potential $ V(r) $,  Eq.~(\ref{eqradpsi}) becomes
\begin{eqnarray}
\frac{d^2\chi(r)}{dr^2}+\left[\omega^2+ {\cal U}_{eff}(r)\right] \chi(r) = 0,
\end{eqnarray}
for the following effective potential
\begin{eqnarray}
\label{pot1}
{\cal U}_{eff}(r)= \frac{4(m+P^2)\omega^2}{r}+\frac{12\ell^2}{r^2}
+\frac{4m^2P^2\omega^2}{r^2} \left[1+3P^2-\frac{3a^2_0}{16m^4P^2}\right]+\cdots,
\end{eqnarray}
 with $\ell^2$ defined as
\begin{eqnarray}
\label{ell}
\ell^2\equiv-\frac{(l^2+l)}{12}+m^2\omega^2\left( 1+\frac{a^2_0}{16m^4} \right)\left(1+P^2\right),
\end{eqnarray}
because of the modification of the term $1/r^2$~\cite{Anacleto:2017kmg,Anacleto:2019tdj}.
We can observe that the suitable asymptotic behavior is satisfied, that is for $ r\rightarrow\infty $ we have $ {\cal U}_{eff}(r) \rightarrow 0 $.

Next we will apply the following approximated formula
\begin{eqnarray}
\label{formapprox}
\delta_l\approx 2(l-\ell)=2\left(l - \sqrt{-\frac{(l^2+l)}{12}
+m^2\omega^2\left( 1+\frac{a^2_0}{16m^4} \right)\left(1+P^2\right)}\right), 
\end{eqnarray}
and taking the limit $ l\rightarrow 0 $, we get the phase shift $ \delta_{l} $ given by
\begin{eqnarray}
\label{phase2}
\delta_l=-{2m\omega}\left( 1+\frac{a^2_0}{16m^4} \right)^{1/2}\left(1+P^2\right)^{1/2}+{\cal O}(l).
\end{eqnarray}
Therefore, knowing $ \delta_{l} $, we can now determine the differential scattering cross section and absorption.
In order to obtain the differential scattering cross section  we will consider the following expression~\cite{Yennie1954,Cotaescu:2014jca}
\begin{eqnarray}
\label{espalh}
\dfrac{d\sigma}{d\theta}=\big|f(\theta) \big|^2=\Big| \frac{1}{2i{\omega}}\sum_{l=0}^{\infty}(2l+1)\left(e^{2i\delta_l} -1 \right)
\frac{P_{l}\cos\theta}{1-\cos\theta}\Big|^2.
\end{eqnarray}
Now considering the small angle limit the above equation is rewritten as
\begin{eqnarray}
\label{espalh2}
\frac{d\sigma}{d\theta}&=&\frac{4}{\omega^2\theta^4}\Big|\sum_{l=0}^{\infty}(2l+1)\sin(\delta_{l})
{P_{l}\cos\theta}\Big|^2,
\\
&=&\frac{16m^2}{\theta^4}\left( 1+\frac{a^2_0}{16m^4} \right)\left( 1+P^2\right)\Big|\sum_{l=0}^{\infty}(2l+1)
{P_{l}\cos\theta}\Big|^2.
\label{espalh3}
\end{eqnarray}
Therefore, equation (\ref{espalh3}) for $ l=0 $ becomes
\begin{eqnarray}
\label{diffcsect}
\frac{d\sigma}{d\theta}\Big |^{\mathrm{l f}}_{\omega\rightarrow 0}
=\frac{16m^2}{\theta^4}\left( 1+\frac{a^2_0}{16m^4} \right)\left( 1+P^2\right)+\cdots.
\end{eqnarray}
For $ P=0 $ and $ a_0=0 $ we obtain the result for the Schwarzschild black hole case.
Thus, we find that the differential scattering cross section of the self-dual black hole increases when the parameters $P$ and/or $a_0$ increase. By comparing with the result of the Schwarzschild black hole the dominant term is modified by the polymeric  parameter $ P $ and parameter $ a_0 $. 
At the limit of $ m\rightarrow 0$ the dominant term of equation (\ref{diffcsect}) becomes nonzero and is given by
\begin{eqnarray}
\label{diffcsect2}
\frac{d\sigma}{d\theta}\Big |^{\mathrm{l f}}_{m\rightarrow 0}
\approx\frac{a^2_0\left( 1+P^2\right)}{\theta^4 m^2}
=\frac{2A^2_{min}\left( 1+P^2\right)}{\theta^4 A_{schwbh}},
\end{eqnarray}
where $ {A}_{schwbh} =4\pi r^2_{+}=16\pi{m^2}$ is the area of the Schwarzschild black hole and $ A_{min} $ is the minimal value of area in loop quantum gravity. 
Therefore, at this limit the differential cross section is directly proportional to the minimum area 
$ A_{min}=8\pi a_0 $ and inversely proportional to the area $ A_{schwbh} $ of the Schwarzschild black hole.

Now we will determine the absorption cross section for a self-dual black hole at the low frequency  limit.
Hence the total absorption cross section can be found through the following formula:
\begin{eqnarray}
\sigma_{abs}
=\frac{\pi}{\omega^2}\sum_{l=0}^{\infty}(2l+1)\Big(\big|1-e^{2i\delta_l}\big|^2\Big)
=\frac{4\pi}{\omega^2}\sum_{l=0}^{\infty}(2l+1)\sin^2(\delta_{l}).
\end{eqnarray}
Next we apply the low energy limit by taking $ \omega\rightarrow 0 $. In this case with $ \delta_l $ given by (\ref{phase2}) the absorption for $ l = 0 $ reads
\begin{eqnarray}
\label{abs1}
\sigma_{abs}^{\mathrm{l f}}
&=& 16\pi{m^2}\left( 1+\frac{a^2_0}{16m^4} \right)\left( 1+P^2\right)
={A}_{schwbh}\left( 1+\frac{16\pi^2 a^2_0}{A_{schwbh}^2} \right)\left( 1+P^2\right).
\end{eqnarray}
By making $P = a_0 = 0$ the result for the absorption of the Schwarzschild black hole is recovered.
Note that when the parameters $P$ and/or $a_0$ increase, the absorption cross section value increases.
Furthermore, we can observe that at the zero mass limit the absorption is nonzero due to the contribution of the minimum area $ a_0 $. So for $ m \rightarrow 0 $ equation (\ref{abs1}) becomes
\begin{eqnarray}
\label{abs1a}
\sigma_{abs}^{\mathrm{l f}}
&\approx &\frac{\pi a^2_0}{m^2}\left( 1+P^2\right)= \frac{16\pi^2 a^2_0}{A_{schwbh}}\left( 1+P^2\right)=\frac{A^2_{min}}{4 A_{schwbh}}\left( 1+P^2\right).
\end{eqnarray}
It is interesting to note that, contrarily to the usual case, i.e., the Schwarzschild black hole, the 
differential scattering/absorption cross section of a self-dual black hole is different from zero as the mass goes to zero. 
{Thus, at this limit the result for absorption cross section increases with the minimum area $ A_{min} $ and decreases with  the area of the Schwarzschild black hole $ A_{schwbh} $. 
It is worth mentioning that in~\cite{Anacleto:2018acl} we have considered an extended Abelian Higgs model with higher order derivatives terms from which an acoustic metric in 2+1 dimensions has been obtained. The absorption in such a context follows an analogous behavior, i.e.,
\begin{eqnarray}
\label{absmhd}
\sigma_{abs}^{\mathrm{l f}}
&\approx &\frac{2\pi\lambda^2C^2}{D}\left( 1+2\lambda^2\right),
\end{eqnarray}
in the limit of $D\to0$. Here $ \lambda $ is the parameter related to the strength of higher order derivatives. $ C $ and $D$ are the  circulation 
and draining parameters. 
This seems to reveal an interesting correspondence between the aforementioned acoustic model in 2 + 1 dimensions and the present gravitational model in 3 + 1 dimensions.
}

Equation (\ref{abs1}) can also be rewritten in terms of the event horizon area of the self-dual black hole as follows
\begin{eqnarray}
\label{abs2}
\sigma_{abs}^{\mathrm{l f}}
\approx4\pi\left[4m^2\left( 1+\frac{a^2_0}{16m^4} \right)\left( 1+P^2\right)\right]=4\pi{\rho_{h}^2}\left( 1+P^2\right)
={\cal A}_{sdbh}\left( 1+P^2\right),
\end{eqnarray}
where 
\begin{eqnarray}
\rho_h=\rho(r_{+})=\sqrt{4m^2+\frac{a^2_0}{4m^2}}=\left[4m^2\left(1+\frac{a^2_0}{16m^2}  \right)\right]^{1/2},
\end{eqnarray}
is the event horizon and 
$ {\cal A}_{sdbh}=4\pi{\rho_{h}^2} $ is the area of the event horizon of the self-dual black hole.

\subsection{Numerical analyses}
Here we present the numerical results that were obtained by numerically solving the radial equation (\ref{eqrad}). 
For this purpose we have adopted the numerical procedure performed in~\cite{Dolan:2012yc}.
The table below shows the results for some values of $ m $, setting the value of $ a_ {0} = \sqrt {3} / 2 $ and $ P = 0.14725 $. All the results are divided by $ \pi $.
\begin{center}
\begin{tabular}{ c||c|c|c  }
 \hline
 $m$     & Equation \eqref{abs1} & Equation \eqref{abs1a} & Numerical result\\
 \hline
 1    & 17.1132  & 0.76626 &  17.1197\\
 0.5  & 7.15178  & 3.05605 &  7.15201 \\
 0.3  & 9.98525  & 8.51403 &  9.98860\\
 0.2  & 19.8104  & 19.1566 &  19.8171\\
 0.1  & 76.7897  & 76.6263 &  76.8165\\
 0.05 & 306.546  & 306.505 &  306.546\\
 0.01 & 7662.63  & 7662.63 &  7662.26\\
  \hline
\end{tabular}
\end{center}
Note that the results of the equations (\ref {abs1}) and (\ref{abs1a}) are the same for small $ m $ as it should be, since equation (\ref {abs1a}) is valid only in the limit of very small values of $ m $. As can be seen in the table, we get a good agreement between them for $ m = 0.01$.

\begin{figure}[htbh]
 \centering
{\includegraphics[scale=0.5]{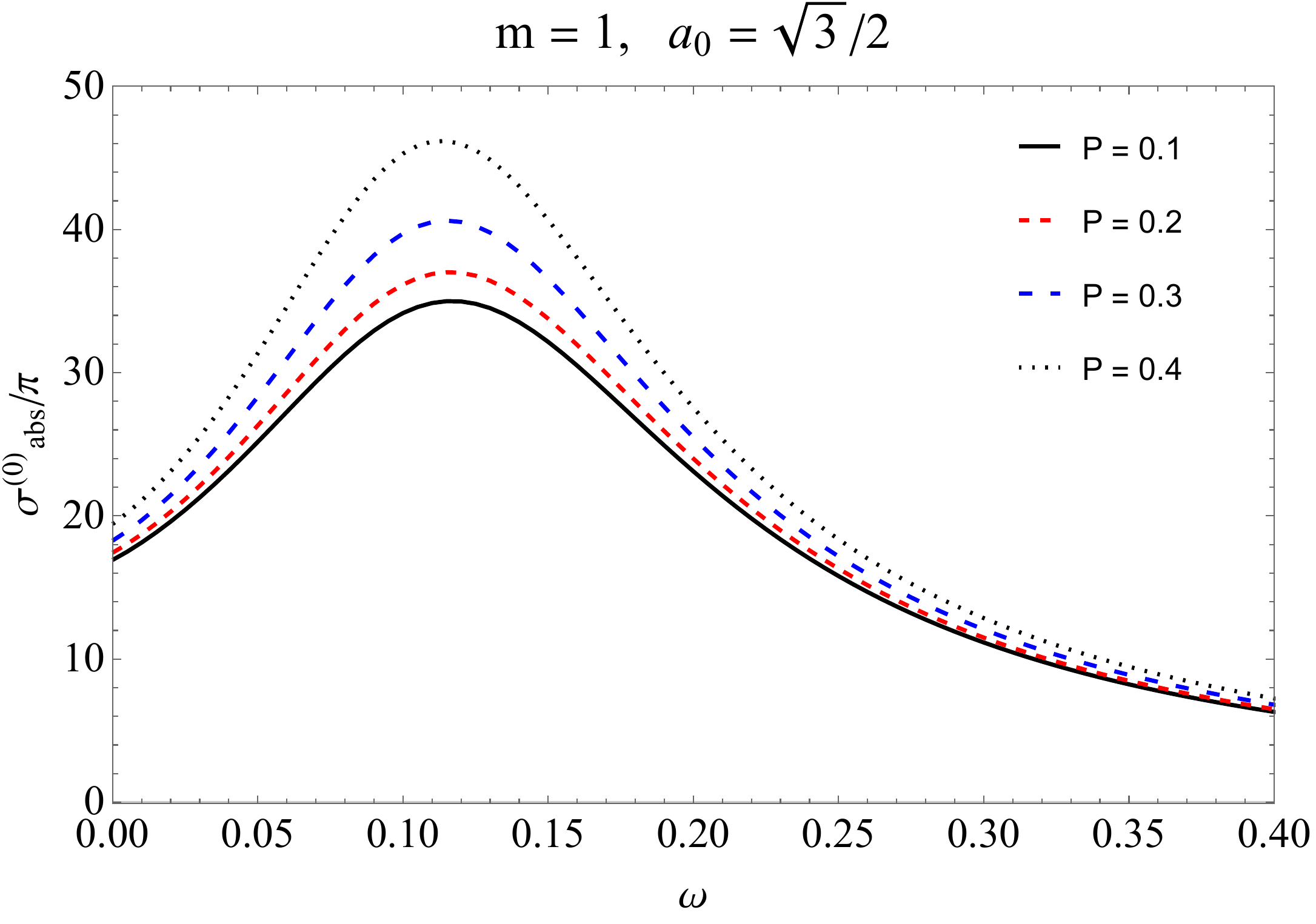}}
 \caption{Partial absorption cross section for $ l=0 $, with $ a_0={\sqrt{3}}/{2} $, $m=1$ and $ P=0.1, 0.2, 0.3, 0.4 $. }
\label{abslo}
\end{figure}

\begin{figure}[htbh]
 \centering
{\includegraphics[scale=0.5]{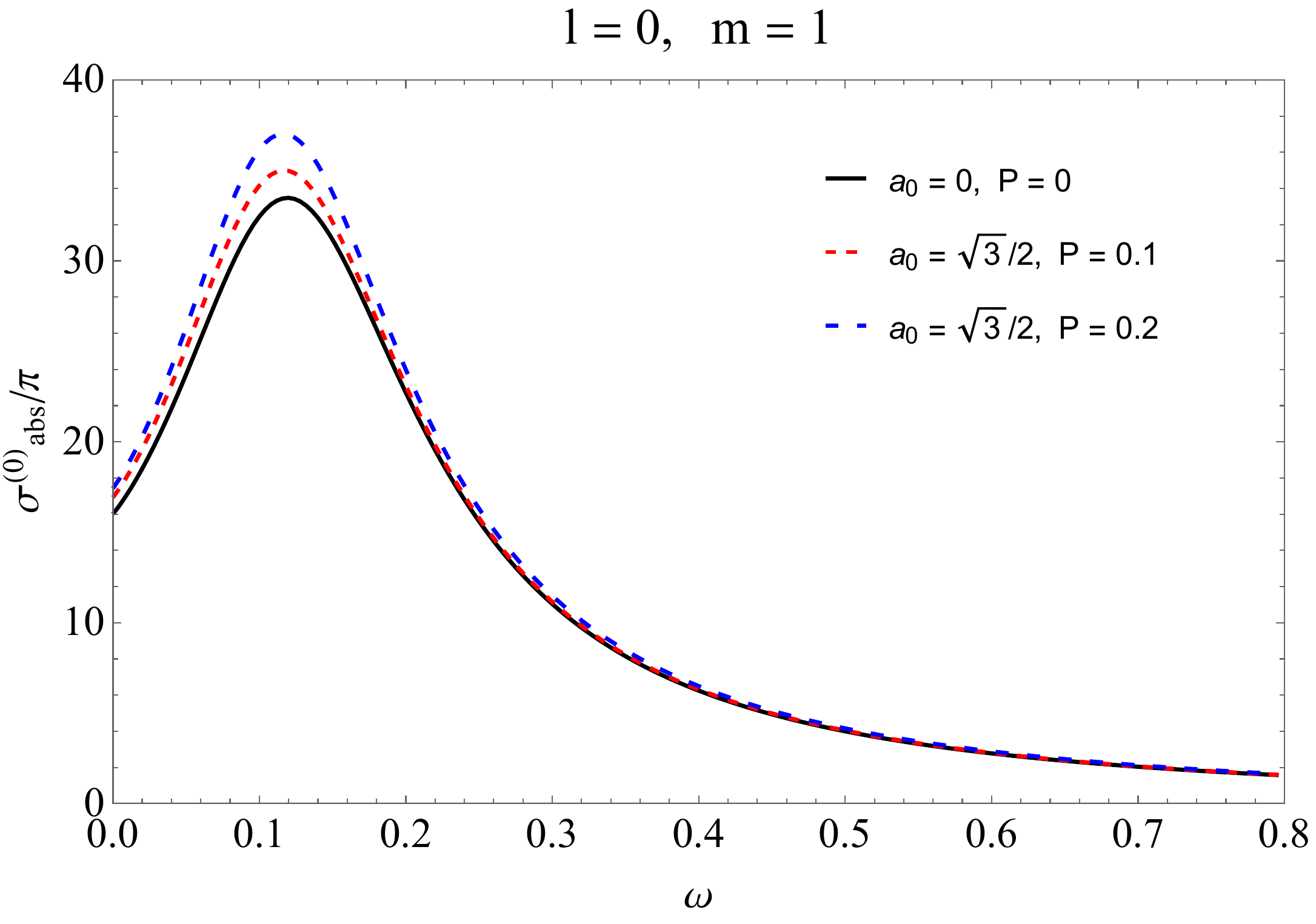}}
 \caption{Partial absorption cross section for $ l=0 $, with $ a_0=0, {\sqrt{3}}/{2} $, $m=1$ and $ P=0, 0.1, 0.2$. }
\label{abslosch}
\end{figure}

\begin{figure}[htbh]
 \centering
{\includegraphics[scale=0.5]{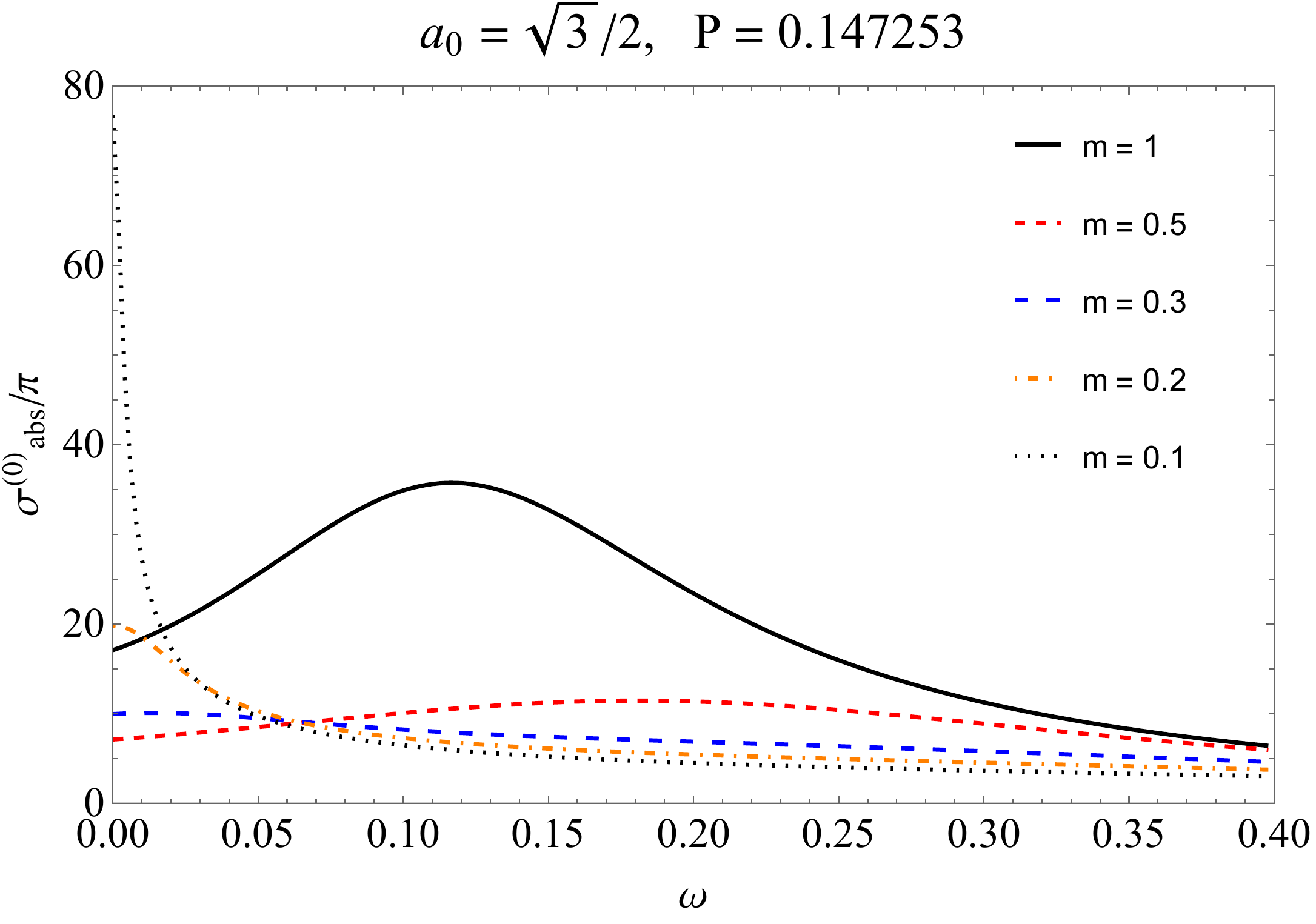}}
 \caption{Partial absorption cross section for $ l=0 $, with $ a_0={\sqrt{3}}/{2} $, $ P=0.147253 $ and $ m=1, 0.5, 0.3, 0.2, 0.1.  $ }
\label{absloa}
\end{figure}

\begin{figure}[htbh]
 \centering
{\includegraphics[scale=0.25]{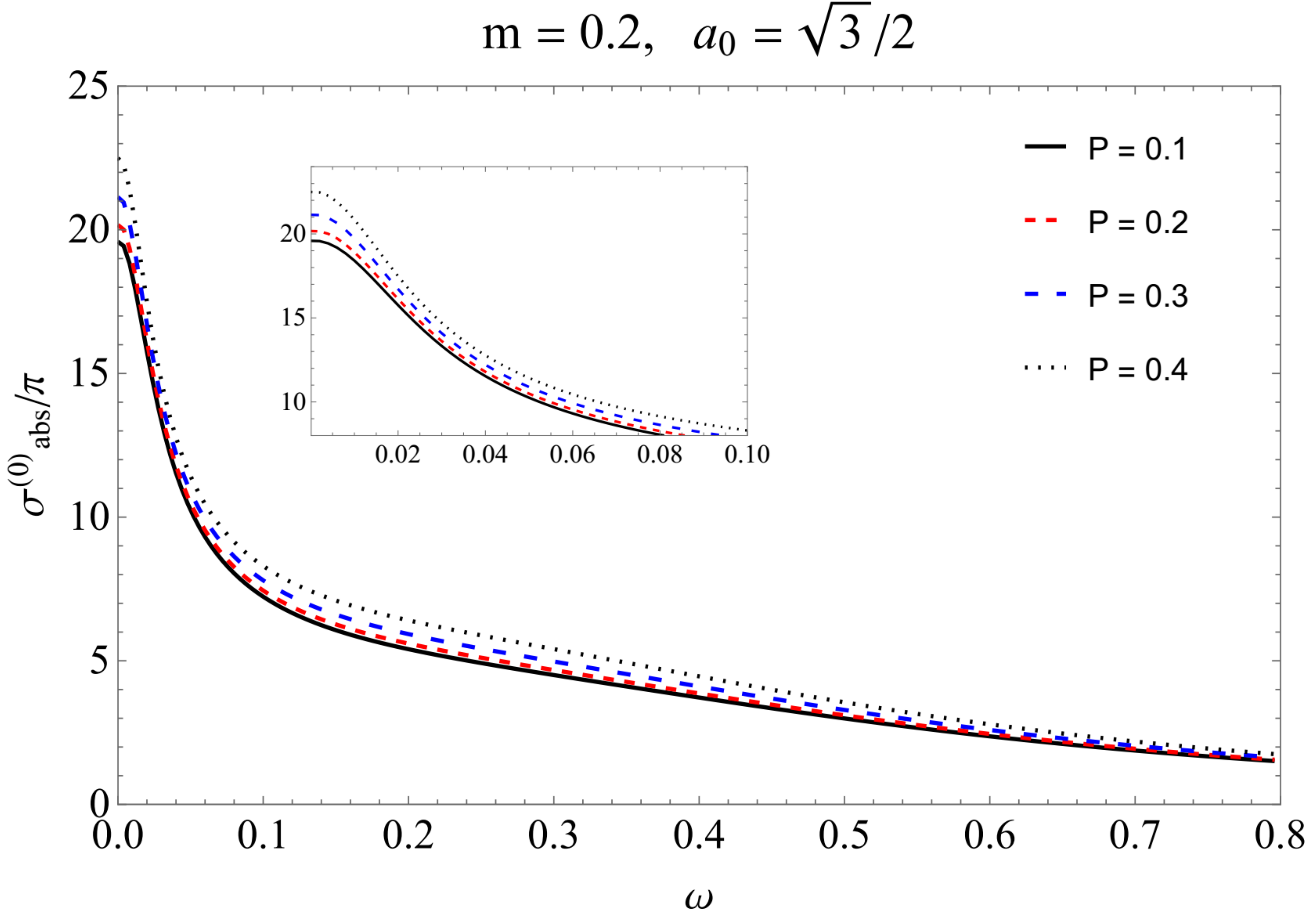}}
 \caption{Partial absorption cross section for $ l=0 $, with $ a_0={\sqrt{3}}/{2} $, $m=0.2$ and $ P=0.1, 0.2, 0.3, 0.4 $. }
\label{abslob}
\end{figure}

\begin{figure}[htbh]
 \centering
{\includegraphics[scale=0.5]{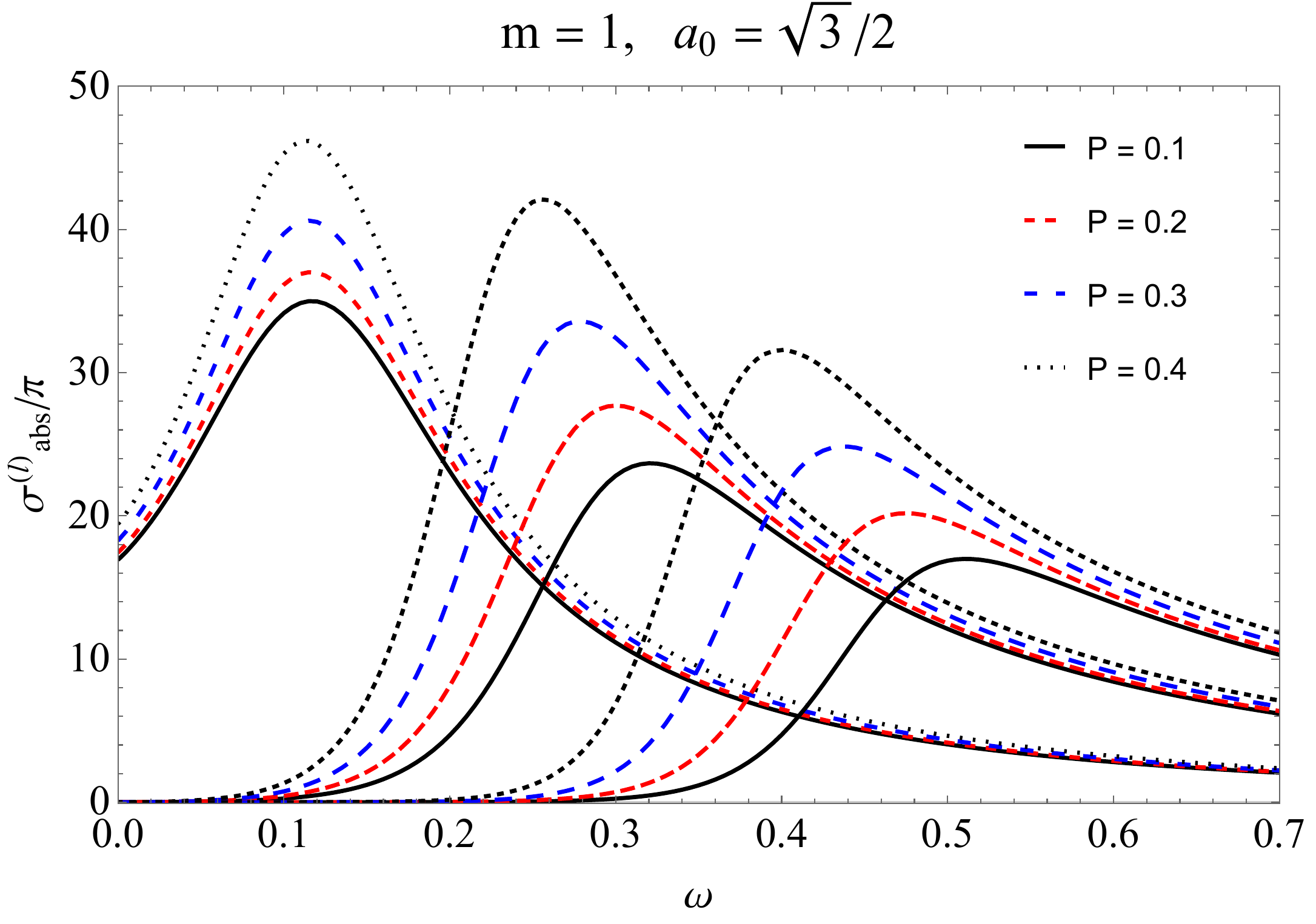}}
 \caption{Partial absorption cross section for modes $ l=0,1,2$, with $ a_0={\sqrt{3}}/{2} $, $m=1$ and $ P=0.1, 0.2, 0.3, 0.4 $.}
\label{absld}
\end{figure}

We can summarize the results as follows. In Fig.~\ref{abslo}, we have plotted the partial absorption for mode $ l = 0 $ with $ m = 1 $, $ a_0 = {\sqrt{3}}/{2} $ by adopting the following values for the polymeric parameter: $ P = 0.1, 0.2, 0.3, 0.4 $.
Analyzing the curves, we find that the absorption amplitude of the self-dual black hole is increased as we vary the $ P $ polymeric parameter.
A comparison between the absorption of the Schwarzschild black hole with that of the self-dual black hole is shown in Fig.~\ref{abslosch}.
The graph for the absorption of the former corresponds to the case where $ a_0=0 $ and $ P = 0 $.
Note that by setting $ a_0 = \sqrt{3}/2 $ and varying $ P $, the partial absorption for the $ l = 0 $ mode has its amplitude increased compared to that of the Schwarzschild black hole.
In Fig.~\ref{absloa} we plot the partial absorption for $ l = 0 $ mode by setting the values of $ a_0=\sqrt{3}/2 $ and $ P=0.147253 $.
We can observe that when we decrease the mass value the absorption amplitude does not tend to zero, as seen earlier from the analytical result shown in equation (\ref{abs1a}).
In Fig.~\ref{abslob} we plot the partial absorption for $ l = 0 $ mode by setting the values of $ a_0=\sqrt{3}/2 $ and  small mass $m=0.2$.
We observe that by varying the $ P $ parameter the absorption amplitude still increases.
The partial absorption cross section graphs for $l= 0,1,2$ modes are shown in Fig.~\ref{absld}.
We have assigned values for the $ P $ polymeric parameter but keeping the values of $ a_0= \sqrt{3}/2 $ and $ m = 1 $ fixed.
Then, as $ P $ increases, partial absorption increases in amplitude.
For modes $ l = 1 $ and $ l = 2 $, note that the curves start from zero, increase in amplitude reaching a maximum value and then decrease in amplitude with the increasing of the frequency $\omega $. 
It is also observed that the maximum partial absorption decreases as the $l$ mode increases.

\section{Conclusions}
\label{conc}
In the present study, we investigated the process of massless scalar wave scattering due to a self-dual black hole through the partial wave method.
We have computed the phase shift analytically at the low energy limit, and then have shown that the dominant contribution at the small angle limit of the differential scattering cross section is modified due to the parameters $ a_0 $ (minimum area) and $ P$ (polymeric function).
We have also found that the result for the absorption cross section is given by the event horizon area of the self-dual black hole at the low frequency limit. 
And mainly, contrarily to the Schwarzschild black hole, the differential scattering/absorption cross section of a self-dual black hole is nonzero at the zero mass limit.
Thus, at the limit of $ m\rightarrow 0 $ the absorption cross section presents a dominant contribution that is inversely proportional to the mass squared, i.e., $ \sigma_{abs}^{\mathrm{l f}}\approx {\pi a^2_0}\left( 1+P^2\right) /m^2$.
In addition, we have verified these results by numerically solving the radial equation for arbitrary frequencies. Finally, we have obtained that the partial absorption amplitude of the self-dual black hole increases its value as we increase the values of the $ P $ parameter in several scenarios. Further investigations, such as exploring correspondence with analog models, may reveal new physics and should be addressed elsewhere.

\acknowledgments
We would like to thank CNPq, CAPES and CNPq/PRONEX/FAPESQ-PB (Grant no. 165/2018), for partial financial support. MAA, FAB and EP acknowledge support from CNPq (Grant nos. 306962/2018-7, 312104/2018-9, 304852/2017-1). We also thank C.A.S. Silva and M.B. Cruz for discussions.

\end{document}